\documentclass[paper,floatfix]{revtex4-1}

\usepackage{amsfonts,amssymb,amsmath,array}
\usepackage{graphicx}

\begin{document}

\title{Stochastic thermodynamics of a piezoelectric energy harvester model}

\author{L. Costanzo}
\affiliation{Department of Engineering, University of Campania ``Luigi Vanvitelli'',  81031, Aversa (Caserta), Italy}

\author{A. Lo Schiavo}
\affiliation{Department of Engineering, University of Campania ``Luigi Vanvitelli'',  81031, Aversa (Caserta), Italy}

\author{A. Sarracino}
\affiliation{Department of Engineering, University of Campania ``Luigi Vanvitelli'',  81031, Aversa (Caserta), Italy}

\author{M. Vitelli}
\affiliation{Department of Engineering, University of Campania ``Luigi Vanvitelli'',  81031, Aversa (Caserta), Italy}

\begin{abstract}We experimentally study  a piezoelectric energy harvester
  driven by broadband random vibrations.  We show that a linear model,
  consisting of an underdamped Langevin equation for the dynamics of
  the tip mass, electromechanically coupled with a capacitor and a
  load resistor, can~accurately describe the experimental data.  In
  particular, the theoretical model allows us to define fluctuating
  currents and to study the stochastic thermodynamics of the system, with
  focus on the distribution of the extracted work over different time
  intervals.  Our analytical and numerical analysis of the linear
  model is succesfully compared to the experiments.
\end{abstract}

\maketitle

\section{Introduction}

From microscopic organisms in the biosphere, life in general and human
activities in particular critically depend on the conversion of
different forms of energy into useful work.  Harvesting energy from
the environment is therefore a central task in many applications,
where random fluctuations possibly arising from disparate sources at
different scales, from~the microscopic thermal Brownian motion in a
fluid, to~the macroscopic vibrations in means of transport, can be
converted into~work.

The well established rules of thermodynamics for macroscopic systems
become blurred when fluctuations are relevant and have to be taken
into account~\cite{sientropy}. From~a theoretical perspective, the~study of fluctuations is addressed within the theory of stochastic
thermodynamics, where the standard concepts of energy, heat, work and
entropy, are extended to non-equilibrium systems, driven by external
forces or coupled to different reservoirs. In~this framework, the~interest is focused on the fluctuations of the above quantities
defined along a single trajectory in the stochastic motion of the
system and on their probability distributions. Indeed, general
relations have pushed the range of validity of standard thermodynamics
into the realm of non-equilibrium regimes~\cite{seifertrev,BPRV08,jepps}:
from the Fluctuation Relations~\cite{GC,ECM,ESR2,LRladek,CM99} and the
generalized fluctuation-dissipation relations~\cite{BM13,temprev}, to~the
general results ruling work and heat exchanged in non-equilibrium
transformations, such as the Jarzinski relation~\cite{CJ}, the~Crooks
fluctuation theorem~\cite{GK}, or~the Hatano--Sasa
relation~\cite{HS01}.  Very recently, thermodynamic uncertainty
relations bounding the signal to noise ratio of a measured current
have been also discovered~\cite{PhysRevLett.114.158101}.  In~particular, the~study of work and heat fluctuations has been the
object of focus in several systems, such as overdamped linear Langevin
Equation \cite{F02}, particle diffusion in time-dependent
potentials~\cite{Maesrecent,PhysRevE.83.061145,PhysRevE.88.062102,Holubec_2015,XIAO2019161,paraguass},
Brownian particles driven by correlated
forces~\cite{PhysRevE.90.052116}, general thermal
systems~\cite{PhysRevE.101.030101}, asymmetric
processes~\cite{Albay_2020}, underdamped Langevin
Equation \cite{Rosinberg_2016}, or~in transient relaxation
dynamics~\cite{CSZ}.  The~interest in these quantities is motivated by
the search for optimization protocols in models of stochastic engines
or, from~a more theoretical perspective, by~the general symmetry
properties or by singular behaviors that work and heat distributions
can show~\cite{corberi}. Experimental studies confirming theoretical
predictions have been reported for instance
in~\cite{ciliberto-1,exptss,ciliberto-2,ciliberto-3}.

Energy harvesting model systems represent an interesting context where
the concepts of stochastic thermodynamics can be applied, due to the
fundamental relevance of random fluctuations. Very well studied
examples are the Brownian (or molecular) motors~\cite{R02}, also known
as ratchet models, where a probe is in contact with a thermal bath and
the presence of a spatial asymmetry coupled to some non-equilibrium
source allows to rectify the motion of the probe, with~extraction of
useful work. These systems have been studied theoretically and
experimentally for instance in the context of granular
media~\cite{puglisi13,GSPP13}, where the dissipative interactions
among grains induce the non-equilibrium behavior, or~in biological
motors~\cite{Leonardo9541}, where active internal forces are at~play.

\begin{figure}[ht!]
  \center
  \includegraphics[width=0.45\textwidth]{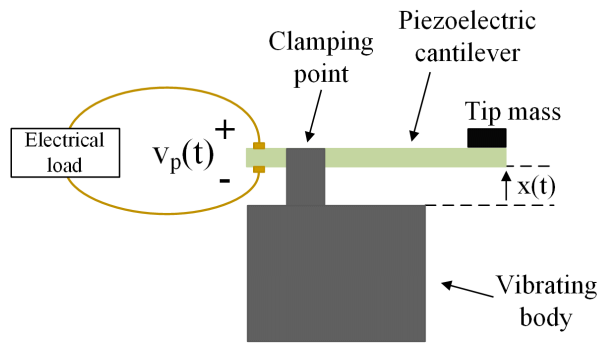}
     \caption{Schematic representation of a cantilever structure with piezoelectric~harvester.}
  \label{fig_scheme}
\end{figure}

More application studies have been carried on in different kinds of
energy harvesters, that are based on the piezoelectric properties of
some materials. In~this case, macroscopic vibrations of the system,
as~for instance in a car or in a train, can induce small currents,
from~which an output power can be extracted to feed sensors or small
electrical devices~\cite{kim}.  Typically, piezoelectric harvesters
are employed in resonant cantilever structures (see
Figure~\ref{fig_scheme}). The~mechanical to electrical energy
conversion mechanism is based on the piezoelectric effect that is the
ability of some materials (notably crystals and certain ceramics) to
generate an electric voltage in response to an applied
mechanical~stress.

Due to their resonant nature, piezoelectric harvesters are typically
studied in steady state sinusoidal conditions at frequencies belonging
to their resonance band~\cite{H1,H2}. In~particular, the~main focus is
on their energetic performance, that is on the mechanical and power
electronic architectures and on the control techniques leading to the
maximization of the extracted power~\cite{H3,H4}. Less attention has
been devoted to the case of resonant piezoelectric harvesters excited
by non-sinusoidal vibrations or solicited by white noise
vibrations~\cite{halv,A,C,D}. In~particular, the~theoretical analysis
of~\cite{halv} provided a stochastic description of the output power
from resonant energy harvesters driven by broadband vibrations and
output power dependence on signal bandwidth was considered. Instead,
Ref.~\cite{C} proposed a methodology for the probabilistic analysis of
a cantilever piezoelectric harvester under white Gaussian noise,
without experimental validation.  In \cite{D}, an experimental analysis
on piezoelectric harvesters is carried out in the presence of
harmonic, random and sine on random vibrations with particular
reference to the electrical power extraction. However, in~all previous
studies on these energy harvesters no attention was devoted to the
analysis of the fluctuations and distributions of relevant quantities
such as the extracted power, in~the general framework of stochastic
thermodynamics of non-equilibrium~systems.

Here we consider a typical piezoelectric harvester in a resonant
cantilever structure driven by random broadband vibrations. Despite
the several nonlinearities present in the experimental system, we show
that a linear model with effective parameters can well reproduce the
observed dynamics. In~particular, we consider a mass in the presence
of a harmonic potential, in~contact with a source of white noise. The~mass is also electromechanically coupled with a capacitor, which allow
for power extraction through a load resistance.  First, we show that
the characteristic response curve, output power vs. load resistance,
obtained from experiments is very well fitted by the analytical
formula derived for the theoretical model. Then, we define a
fluctuating work along a system trajectory, according to the standard
approach of stochastic thermodynamics, and~we focus on the study of
the work fluctuations. We find that also the distributions of the work
measured experimentally over different time intervals, are in very
good agreement with those computed in numerical simulations of the
linear model, using effective~parameters.

Our study presents an experimental characterization of the
distributions of integrated currents (output power) for a system that
is used in applications as a valid energy harvester device. Moreover,
our analysis shows that a simplified linear model, which allows for
analytical computations, is able to accurately reproduce the
experimental results, even at the fine level of~fluctuations.

\section{Experimental~Setup}

\begin{figure}[ht!]
  \center
  \includegraphics[width=0.45\textwidth]{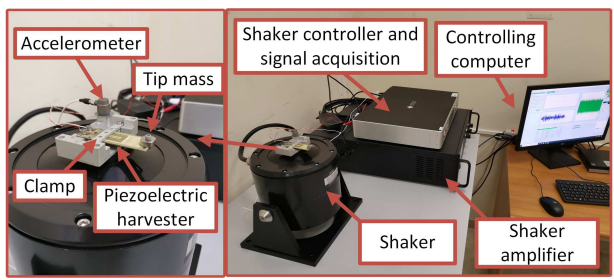}
  \caption{Picture of the experimental~setup.}
  \label{picture}
\end{figure}

The schematic representation of the experimental setup for the
piezoelectric harvester is shown in Figure~\ref{picture}. It consists
of a cantilever structure with a tip mass, whose displacement in time
$x(t)$ due to the vibrations provided by the shaker, induces a voltage
$v_p(t)$ across the electrical~load.

In all the experimental tests that we have carried out, we have used
the commercial piezoelectric harvester MIDE PPA-4011 loaded by
different electrical load resistances. This product incorporates four
piezoelectric wafers resulting in significant performance improvements
with respect to other models by MIDE. The~harvester has been mounted
on a shaker by using a support providing the possibility of different
clamping positions. The~addition of tip masses in order to define
mechanical properties of the resonant structure is also
possible. A~picture of the whole experimental setup is shown in
Figure~\ref{picture}. The~shaker VT-500 by Sentek (500 N rated force
and 450 m/s$^2$ maximum acceleration) has been used to get the desired
input vibrations. The~controller Spider 81 allowed application of the
desired voltage signal to the shaker amplifier and to carry out the
recording, by~means of its built-in acquisition board, of~the voltage
across the load resistance. An~accelerometer 3055D2 by Dytran
(sensitivity 100 mV/g on the range 50 g) has been used to monitor the
applied acceleration in order to implement a closed-loop feedback
vibration~control.

\begin{figure}[ht!]
  \includegraphics[width=0.5\textwidth]{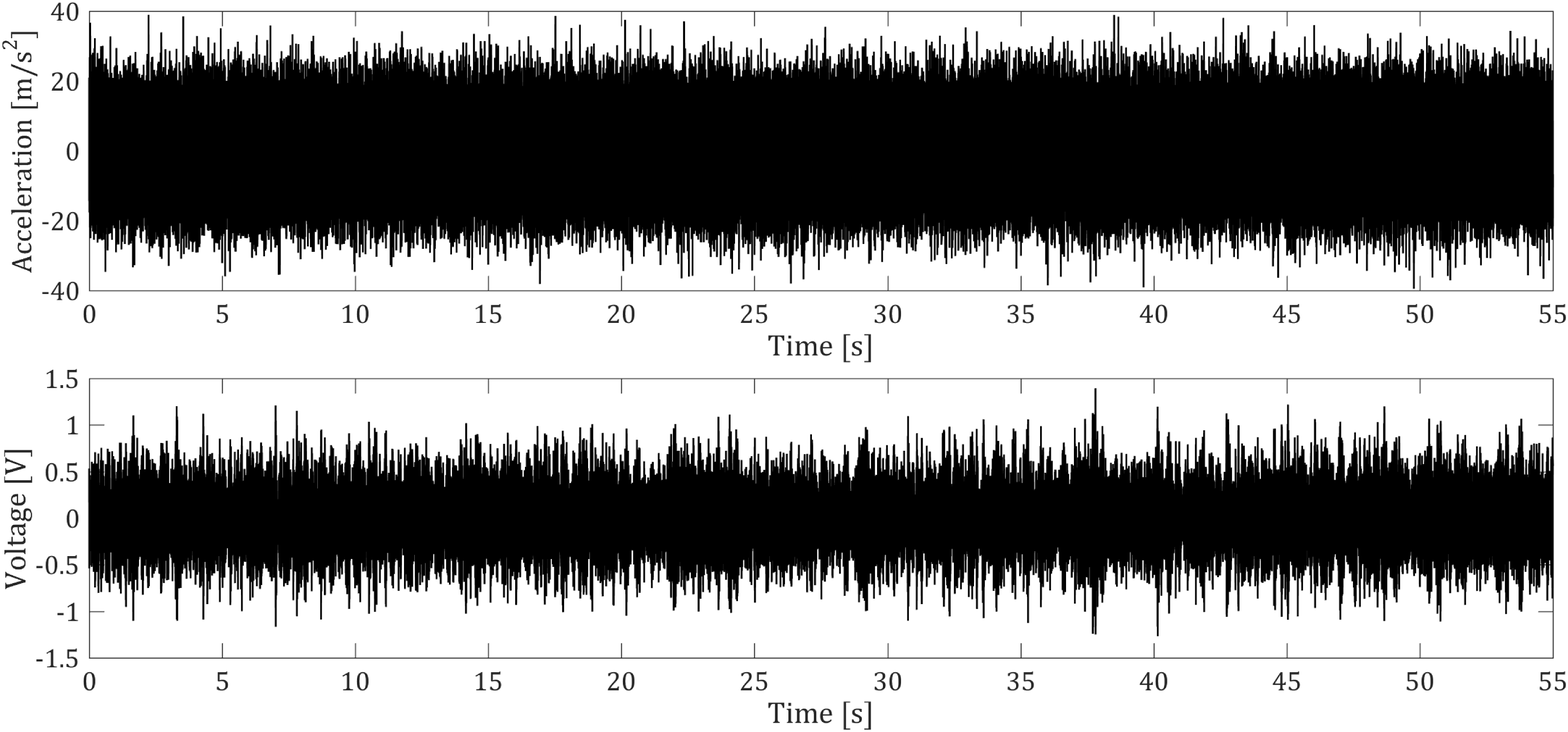}
  \includegraphics[width=0.5\textwidth]{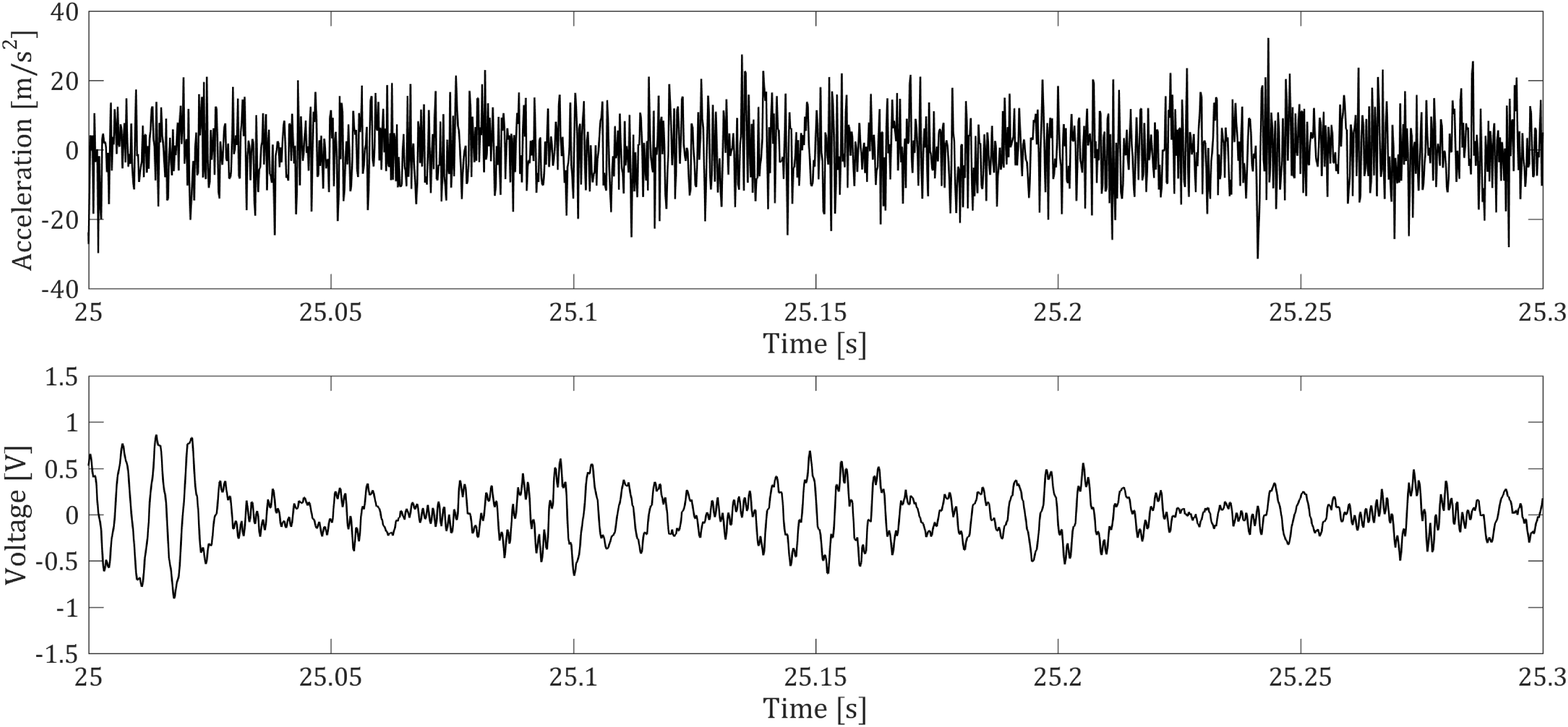}
  \caption{Top: Input white noise acceleration and voltage across a
    2200 $\Omega$ load resistance. Bottom: zoom on a time window of 0.3~s.} 
  \label{wave}
\end{figure}

In order to study the response of the system to broadband vibrations,
we fed the shaker with a Gaussian signal generated with standard
software (MATLAB), with~a sampling rate $f=5$ kHz.  In~Figure~\ref{wave}
typical waveforms of the input acceleration and of the voltage across
the load resistance (for $R=2200~\Omega$) recorded during experimental
tests are~shown.

\begin{figure}[ht!]
  \center
  \includegraphics[width=0.45\textwidth]{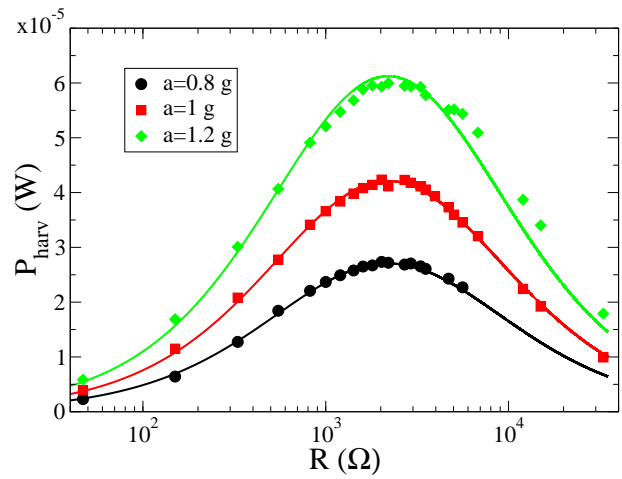}
  \caption{$P_{harv}$ (measured in Watt) as a function of the load
    resistance, for~different values of the input acceleration $a$,
    measured in unit of the gravity acceleration $g$. Symbols are
    experimental data, while lines correspond to the
    formula~(\ref{pharv0}).}
  \label{fig0}
\end{figure}

The first quantity we analyzed is the extracted power $P_{harv}$ that
can be obtained from the average dissipated heat on the load
resistance for unit time, $P_{harv}=\langle v_p^2\rangle/R$.  We show
in Figure~\ref{fig0} the characteristic curves $P_{harv}$ vs. $R$, for~different values of the input acceleration. From~our analysis, we find
an optimal resistance value $R^*\sim 2000~\Omega$, which is independent of
the shaking~amplitude.

\section{Theoretical~Model}

In order to describe the observed experimental results and to extend
the study of the system to fluctuating quantities relevant in the
stochastic thermodynamics framework, we consider the
following linear model
\begin{eqnarray}
  \dot{x}&=&v \label{lang1}\\
  M\dot{v}&=&-K_sx -\gamma v -\theta v_p + M \xi \label{lang2}\\
  C_p\dot{v}_p&=&\theta v -\frac{v_p}{R}, \label{lang3}
\end{eqnarray}
where $\xi$ is white noise with zero mean and correlation $\langle
\xi(t)\xi(t')\rangle = 2D_0\delta(t-t')$.  In the above equations, $x$
represents the displacement of the tip mass $M$, $v$ its velocity,
$\gamma$ the viscous damping due to the air friction, $K_s$ the
stiffness of the cantilever in the harmonic approximation, $v_p$ the
voltage across the load resistance $R$, $C_p$ the effective
capacitance in the circuit, and~$\theta$ the effective
electromechanical coupling factor of the
transducer. In~Equation~(\ref{lang2}) we have neglected the thermal
fluctuations on the tip mass, which are too small to affect
its~motion.

In the system we can identify several characteristic times:
$\tau_1=M/\gamma, \tau_2=\sqrt{M/K_s},$ $\tau_3=C_p\gamma/\theta^2,
\tau_4=C_pR$. $\tau_1$ is the relaxation time of the tip mass due to
the viscous damping, $\tau_2$ is the relaxation time within the
harmonic potential, $\tau_3$ represents the typical timescale of the
coupling between the proof mass and the capacitor, $\tau_4$ is the
characteristic time of the RC circuit. Among~the various
characteristic times, $\tau_3$ is the only one depending on quantities
belonging to both the electrical and the mechanical subdomain of the
whole harvesting system. Hence, its physical meaning is not as
intuitive as in the case of the other characteristic times. In~any
case, in~order to better highlight its role, it is possible to
show~\cite{H4} that there is a link between the amplitude of the speed
of the tip mass and the amplitude of the external acceleration, when
the harvester operates in sinusoidal conditions at the resonance
frequency and in open circuit (no load, $R\to\infty$). In~particular,
one has the relation~\cite{H4}
\begin{equation}
v_{max}= a_{max} \theta \frac{M}{\gamma\sqrt{1+ \frac{1}{(K_s/M)\tau_3^2}}} = a_{max} \theta \frac{\tau_1}{\sqrt{1+ \tau_2^2 /\tau_3^2}}. 
\end{equation}
Therefore, the~speed amplitude, in~the above operating conditions,
depends on all the three characteristic times, $\tau_1$, $\tau_2$,
and~$\tau_3$, that assume finite values. It does not depend on
$\tau_4$, since it is unbounded in open circuit ($R\to\infty$).

The system of Equations~\eqref{lang1}--\eqref{lang3} can be mapped onto a non-Markovian model,
which makes clear how the presence of the coupling between the tip
mass and the capacitor introduces a form of memory in the
dynamics. In~particular, one can rewrite the above equations as a
generalized Langevin equation
\begin{equation}
\dot{v}=-\int_0^t\left[2\frac{\gamma}{M}\delta(t')+\Gamma(t-t')\right]v(t')dt'-\frac{K_s}{M}x+\xi,
  \end{equation}
where the memory kernel $\Gamma(t)$ has the simple exponential form
\begin{equation}
\Gamma(t)=\frac{\theta^2}{C_pM}e^{-t/RC_P} = \frac{1}{\tau_1\tau_3}e^{-t/\tau_4}.
\end{equation}
Let us note that here the friction memory kernel $\Gamma(t)$ is not
associated with any noise term. This puts the system by construction
in a non-equilibrium state, because~the fluctuation-dissipation
relation of the second kind does not~hold.

\subsection{Average~Values}

The static properties of the linear model can be obtained by standard methods~\cite{R89}.
We introduce the column vector $X=(x,v,v_p)^T$ and the coupling matrix
\begin{equation}
A=\left(
\begin{array}{ccc}
 0 & -1 & 0 \\
 \frac{K_s}{M} & \frac{\gamma }{M} & \frac{\theta }{M} \\
 0 & -\frac{\theta }{C_p} & \frac{1}{C_p R}
\end{array}
\right),
\end{equation}
so that the Equations~\eqref{lang1}--\eqref{lang3} can be rewritten in vectorial form as
\begin{equation}
\dot{X}=-A X + \eta,
\end{equation}
where $\eta=(0,\xi,0)^T$.
Defining the covariance matrix $\sigma=\langle X^T X \rangle$ as
\begin{equation}
\sigma=\left(
\begin{array}{ccc}
 \sigma_{xx} & \sigma_{xv} & \sigma_{xv_p} \\
 \sigma_{vx} & \sigma_{vv} & \sigma_{vv_p} \\
 \sigma_{v_px} & \sigma_{v_pv} & \sigma_{v_pv_p}
\end{array}
\right),
\end{equation}
at stationarity one has the constraint
\begin{equation}\label{stationary}  
D=\frac{A\sigma+\sigma A^T}{2},
\end{equation}
where $D$ is the noise matrix
\begin{equation}
D=\left(
\begin{array}{ccc}
 0 & 0 & 0 \\
 0 & D_0 & 0 \\
 0 & 0 & 0
\end{array}
\right).
\end{equation}
From Equation~(\ref{stationary}) one gets the following relations
for the covariance matrix elements
\begin{eqnarray}
 0&=& \langle xv\rangle \\
 0&=&\langle v^2\rangle -\frac{K_s}{M}\langle  x^2\rangle -\frac{\theta}{M}\langle xv_p\rangle \\
  0&=&-\frac{\gamma}{M}\langle v^2\rangle -\frac{\theta}{M}\langle vv_p\rangle + D_0  \\
  0&=&\frac{\theta}{C_p}\langle vv_p\rangle -\frac{1}{C_pR}\langle v_p^2\rangle \label{equal}\\
  0&=& -\left(\frac{\gamma}{M}+\frac{1}{C_pR}\right)\langle vv_p\rangle +\frac{\theta}{C_p}\langle v^2\rangle -\frac{K_s}{M}\langle xv_p\rangle -\frac{\theta}{M}\langle v_p^2\rangle \\
  0&=&\langle vv_p\rangle -\frac{1}{C_pR}\langle xv_p\rangle. 
\end{eqnarray}
The stationary distribution is a multivariate Gaussian
\begin{equation}  
P(x,v,v_p)\sim \exp\left[-\frac{1}{2}\left(\sigma_{xx}^{-1}x^2+\sigma_{vv}^{-1}v^2+\sigma_{v_pv_p}^{-1}v_p^2+2\sigma_{xv}^{-1}xv+2\sigma_{xv_p}^{-1}xv_p+2\sigma_{vv_p}^{-1}vv_p\right)\right],
\end{equation}
where $\sigma^{-1}$ denotes the inverse matrix of $\sigma$.  The~explicit
expressions of the elements of $\sigma$ are reported in~Appendix \ref{ppendixA}.

The average output power is the heat dissipated into the resistance
per unit time
\begin{equation}
P_{harv}=\langle \dot{Q}_{diss} \rangle   =
\frac{1}{R}\langle v_p^2\rangle
\end{equation}
and its explicit expression as a function of the parameters is
\begin{equation}\label{pharv0}  
P_{harv}=\frac{D_0 M^2 R \theta^2}{ M (\gamma + R \theta^2) + 
  C_p R \gamma (C_p K_s R + \gamma + R \theta^2)}.
\end{equation}

\subsection{Fitting the Model to Experimental~Data}

The linear model described by the Equations~\eqref{lang1}--\eqref{lang3}
contains several physical parameters that are directly controlled in
the experiments and others that can be fitted to match the measured
data. In~particular, the~parameter $D_0$ that quantifies the amplitude
of the white noise is related to the acceleration $a$ provided by the
shaker and to the sampling rate $1/\Delta t$ of the input signal,
$D_0=a^2\Delta t/2$, where $\Delta t=1/f=0.0002$ s.  Furthermore, the
parameters $K_s$ and $M$ are related to the characteristic frequency of
the system, which for the experimental apparatus is
$\sqrt{K_s/M}=2\pi\times 140$ Hz. Finally, the~capacitor $C_p$ is
estimated as $C_p$$\sim$490 nF.  The~other parameters can be fitted to
the experimental data through the analytical expression~(\ref{pharv0})
as a function of the load resistance $R$. For~the case \mbox{$a$ = 9.81 m/s$^2$
,} we obtain the following values for the model parameters: $M=0.0083\pm 0.0002$
Kg, \mbox{$\theta=0.0195\pm0.002$ N/V,} $\gamma=0.359\pm0.05$ Kg/s. This set of parameters
is used also for other values of the shaker accelerations used in the
experiments, $a=0.8\times 9.81$ m/s$^2$ and $a=1.2\times 9.81$ m/s$^2$,
providing a very good agreement between analytical predictions and
experimental data, as~shown in Figure~\ref{fig0}.

\subsection{Stochastic~Energetics}

The theoretical model allows us to study fluctuations and distributions
of thermodynamic quantities defined at the level of the single trajectory.
In particular, according to Sekimoto~\cite{Sekimoto2010}, we define
the heat exchanged along a trajectory in a time interval $\tau$ with
the surrounding medium as
\begin{equation}\label{heatex}
\mathcal{Q}_{ex}(\tau)=-\int_0^\tau \gamma v(t)^2dt,
\end{equation}
and the energy fed into the system from the external driving as the
integral of the injected power $P_{inj}=M\xi(t)v(t)$
\begin{equation}
\mathcal{E}_{inj}(\tau)=   M\int_0^\tau  \xi(t)v(t) dt.
\end{equation}

The product in the above equation is meant according to the
Stratonovich prescription. Note that the heat in Equation~(\ref{heatex}) is
released toward the medium.  Using the Langevin
Equation~(\ref{lang2}), we can rewrite these two terms as follows
\begin{eqnarray}
  \mathcal{Q}_{ex}(\tau)+\mathcal{E}_{inj}(\tau)&=&\int_0^\tau\left[-\gamma v(t)^2+ M v(t)\xi(t)\right]dt \nonumber\\
  &=&\int_0^\tau\left[-\gamma v(t)^2+  v(t)\left(M\dot{v}+K_s x +\gamma v +\theta v_p\right)\right]dt \nonumber\\
  &=&\frac{1}{2}M[v(\tau)^2-v(0)^2]+\frac{1}{2}K_s[x(\tau)^2-x(0)^2]+  \theta \int_0^\tau v(t)v_p(t)dt \nonumber\\
  &=&\Delta E + W, \label{1princ}
\end{eqnarray}
where
\begin{equation}
  \Delta E=\frac{1}{2}M[v(\tau)^2-v(0)^2]+\frac{1}{2}K_s[x(\tau)^2-x(0)^2]
\end{equation}
is the mechanical energy variation and
\begin{equation}
  W(\tau)=\theta\int_0^\tau v(t)v_p(t)dt
\end{equation}
can be interpreted as the work performed by the harvester. Equation~(\ref{1princ}) represents
the first principle for stochastic thermodynamic quantities.  The~average output
power in the stationary state is then
\begin{equation}\label{pharv1}
P_{harv}=\langle  \dot{W}\rangle=\theta\langle vv_p\rangle=-\gamma\langle v^2\rangle+M\langle v\xi\rangle,
\end{equation}
where $\langle v\xi\rangle=D_0$ and the last equality follows from Equation~(\ref{1princ}) using the stationary result $\langle \Delta E\rangle=0$.
The output power can be also expressed as the dissipated heat in the
load resistance in the time interval $\tau$
\begin{equation}
Q_{diss}(\tau)=\frac{1}{R}\int_0^\tau  v_p(t)^2 dt, 
\end{equation}
which, exploiting Equation~(\ref{lang3}), can be related to the fluctuating
work $W$ by
\begin{equation}  
W(\tau)=Q_{diss}(\tau)+\frac{C_p}{2}[v_p(\tau)^2-v_p(0)^2].
\end{equation}
This shows that the difference between the fluctuations of $W$ and
$Q_{diss}$ is a term non-extensive in time. In~our system, we have
numerically checked that, for~the studied time intervals, these border
terms can be neglected.  However, more generally, there are systems
where such terms can be relevant for fluctuations, see for
instance~\cite{ZC03,Maesrecent}.

\begin{figure}[ht!]
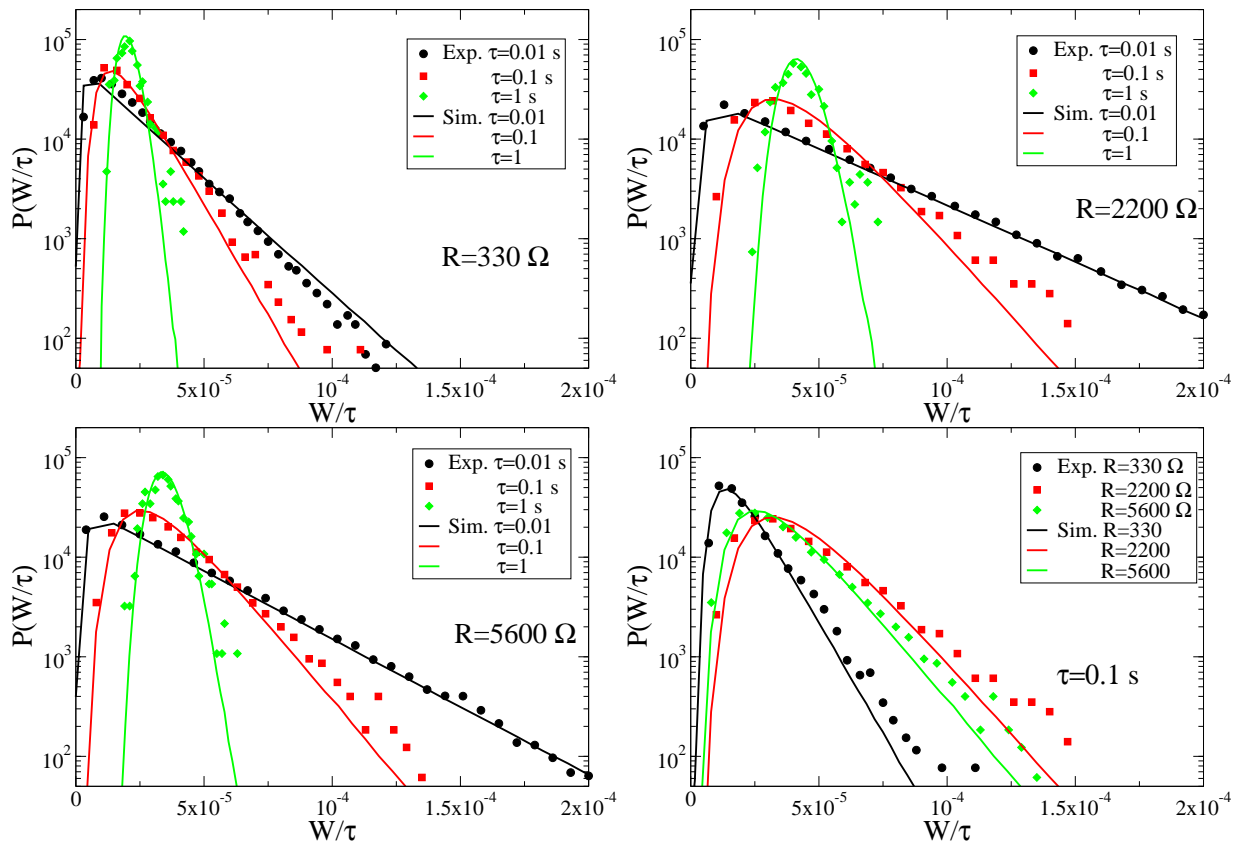

\center
  \includegraphics[width=0.45\textwidth,clip=true]{R330variT.eps}
  \includegraphics[width=0.45\textwidth,clip=true]{R2200variT.eps}
  \includegraphics[width=0.45\textwidth,clip=true]{R5600variT.eps}
  \includegraphics[width=0.45\textwidth,clip=true]{T01variR.eps}
    \caption{Distributions of $W/\tau$ (measured in Watt) for
      different values of the load resistance $R$ and of the time
      $\tau$. Dots represent experimental data and lines numerical
      results. Numerical simulations were obtained integrating the
      Langevin equation with a time step $dt=10^{-6}$, and~averaging
      over $\sim$10$^5$ realizations.}
  \label{fig1}
\end{figure}

We have studied numerically and experimentally the work $W(\tau)$ for
different values of $\tau$. The~experimental evaluation of this
quantity has been obtained integrating the time series of the output
signal for the voltage $v_p(t)$ (the whole recorded time series were 600 s
long). The~results are reported in Figure~\ref{fig1}. First, we stress
the good agreement between the work distributions obtained from
experiments and numerical simulations.  This shows that the linear
model with white noise is able to accurately describe the experimental
system, even at the level of fluctuations, in~the range of explored
parameters.  The~work distributions present a pronounced asymmetry for
small time intervals $\tau$, characterized by an exponential tail for
large values of $W/\tau$, as~also observed for the functional form of
the injected power distribution in the overdamped Langevin equation,
obtained analytically in~\cite{F02}. At~large times, the~distributions
seem to converge towards a Gaussian form, symmetric around the~mean.

\begin{figure}[ht!]
\center
  \includegraphics[width=0.45\textwidth]{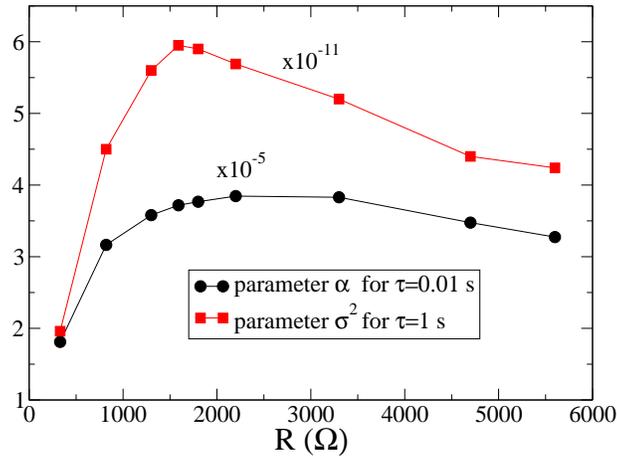}
    \caption{Parameters $\alpha$ and $\sigma^2$ obtained from the fit
      of the functions $f(x)$ and $g(x)$ to the experimental data, for different values of $R$.}
  \label{fig2}
\end{figure}

In order to analyze the behavior of the work distributions measured in
experiments as a function of the load resistance we have fitted the
tail of the curves at short times ($\tau=0.01$ s) with an exponential
function $f(x)\sim\exp(-x/\alpha)$, extracting the parameter $\alpha$. For~the
curves at large times ($\tau=1$ s) we have fitted the data with a
Gaussian function $g(x)\sim\exp(-(x-\mu)^2/2\sigma^2)$ to obtain the variance
$\sigma^2$ as a function of $R$. The~results are shown in
Figure~\ref{fig2}. We observe that both $\alpha$ and $\sigma^2$ have a
non-monotonic behavior, with~a maximum appearing around the value which
maximizes the mean extracted power, showing that fluctuations are
larger in proximity of the optimal working~point.

\section{Conclusions}

We have studied experimentally a piezoelectric energy harvester driven
by random broadband vibrations, focusing on the behavior of the
extracted power as a function of the load resistance and of the
vibration amplitude.  We have shown that a linear model, consisting of
an underdamped Langevin equation for the tip mass coupled with voltage
dynamics reproduces very well the experimental~data.

Moreover, the~theoretical model allowed us to address the issue
related to the behavior of the fluctuations of non-equilibrium
currents, such as the extracted work in a time interval. This analysis
plays a central role in the context of stochastic thermodynamics,
where the properties of the system are scrutinized at the level of
single trajectories. The~comparison between the results of numerical
simulations of the model and the experimental data showed that the
linear system of equations provides a good approximation of the real
system, reproducing the same behavior of the work distributions as a
function of the~parameters.

Our findings represent the first experimental study of the work
fluctuations in a piezoelectric energy harvester and show that the
observed behaviors can be consistently rationalized within a simple
model, paving the way to future analyses. In~particular, from~the
theoretical perspective, due to the linear nature of the model, the~analytical computation of the work distribution and its large
deviations function could be obtained with a path integral approach,
as for instance described in~\cite{F02}. Moreover, it could be
interesting to modify the model by adding a noise source also in the
equation for the voltage, obtaining a system of two coupled Langevin
equations, or~considering a bistable potential for the tip mass, as~proposed in~\cite{gammaitoni,PhysRevLett.102.080601}. Analyses of
other quantities such as heat or entropy production could be also
carried out along the same lines.  On~the experimental side, it could
be interesting to perform a similar study of current fluctuations
in a system driven by realistic vibration sources, like cars or trains,
taken from available databases, or~where the simple linear resistance
load is replaced by a diode bridge, as~often considered in
applications.



\section{Appendix A}\label{ppendixA}
Here we report the explicit expressions for the covariance matrix elements
\begin{eqnarray}
\sigma_{xx} &=&\frac{D_0 M^2 (C_p R (\gamma +C_p K_s R)+M)}{K_s \left(\gamma  C_p^2 K_s R^2+\left(\gamma +\theta^2 R\right) (\gamma  C_p R+M)\right)}\\
 \sigma_{xv}&=&0 \\
 \sigma_{xv_p}&=& \frac{C_p D_0 \theta  M^2 R^2}{\gamma  C_p R \left(\gamma +C_p K_s R+\theta^2 R\right)+M \left(\gamma +\theta^2 R\right)} \\
 \sigma_{vv}&=&\frac{D_0 M \left(C_p R \left(\gamma +C_p K_s R+\theta^2 R\right)+M\right)}{\gamma  C_p R \left(\gamma +C_p K_s R+\theta^2 R\right)+M \left(\gamma +\theta^2 R\right)}\\
 \sigma_{vv_p}&=& \frac{D_0 \theta  M^2 R}{\gamma  C_p R \left(\gamma +C_p K_s R+\theta^2 R\right)+M \left(\gamma +\theta^2 R\right)}\\
  \sigma_{v_pv_p}&=&\frac{D_0 \theta^2 M^2 R^2}{\gamma  C_p R \left(\gamma +C_p K_s R+\theta^2 R\right)+M \left(\gamma +\theta^2 R\right)}. 
\end{eqnarray}


\end{document}